\documentclass[12pt]{report}
\usepackage[dvips]{graphicx}
\newcommand{\sptwo}{1.4}

\newcommand{\doublespace}{\edef\baselinestretch{\sptwo}\Large\normalsize}

\begin{document}
\doublespace
\begin{center}
{\bf A note on the quantum-tail effect on fusion reaction rate
}\\
\renewcommand\thefootnote{\fnsymbol{footnote}}
{
Alexander L. Zubarev\footnote{ e-mail: zubareva$@$physics.purdue.edu}}\\
Department of Physics, Purdue University\\
West Lafayette, Indiana  47907\\
\end{center}

\begin{quote}
A study is made of the power-law tail effect in the quantum particle 
distribution over
momentum on the nuclear fusion reactions.
Our results do not support the idea of averaging the fusion reaction
 cross-section over the momentum distribution postulated 
%in [A.N. Starostin and N.L. Alexandrov, Phys. Plasmas {\bf 5}, 2127 (1998); N.L. Alexandrov and A.N. Starostin, Zh. Eksp. Teor. Fiz. {\bf 113}, 1661 (1998) [JETP {\bf 86}, ]903 (1998)]]
and used in many publications.
\end{quote}

\vspace{5mm}

\pagebreak
While in classical statistics all systems of particles in equilibrium have 
single-particle 
 momentum distribution (MD),
 $n(\vec{p})$,
of Maxwell-Boltzmann form,
the MD in quantum statistics has a non-Maxwellian form,
 contains a non-exponential tail [1-4] and
 plays a role central to our 
understanding  of interacting quantum particles systems.
 $n(\vec{p})$ has to satisfy the sum rules
$$
\int n(\vec{p}) d^3p = 1,
\eqno{(1)}
$$
$$
\frac{1}{2 m}\int n(\vec{p}) p^2 d^3 p = T_k, 
\eqno{(2)}
$$
where $T_k$ is the kinetic energy. 

In order for the kinetic energy integral, Eq.(2), to remain finite, the MD 
should decline faster than $p^{-5}$. The large $p$-behavior of $n(\vec{p})$
 has been considered in many papers [1-4]. It was shown that the interaction
 between quantum particles leads to the appearance of a power-law 
tail
in the MD.
 It was found for the first time in Ref.[2] that the zero temperature MD 
for  
an interacting electron gas should go as $1/p^8$ for large $p$.

A general expression for $n(\vec{p})$ can be written as [5]
$$
n(\vec{p}) = \int f_{\gamma}(E,\vec{p})dE,
\eqno{(3)}
$$
where the energy-momentum distribution $f_{\gamma}(E,\vec{p})$
can be represented as [5]
$$
f_{\gamma}(E,\vec{p})=f(E) \delta_{\gamma}(E,\vec{p}),
\eqno{(4)}
$$
 $f(E)$ is the occupation number (Fermi-Dirac, Bose-Einstein or 
Maxwell-Boltzmann), $\delta_{\gamma}(E,\vec{p})$ is the spectral function [5] 
$$
\delta_{\gamma}(E,\vec{p})=\frac {\gamma (E,\vec{p})}{\pi 
[(E-\epsilon_p-\Delta(E,\vec{p}))^2+\gamma^2 (E,\vec{p})]},
\eqno{(5)}
$$ 
$\epsilon_p=p^2/(2 m)$, and $\gamma (E,\vec{p})$ and $\Delta(E,\vec{p})$ are solutions of ``complex set of integral equations" [5].

The finite value of $\gamma (E,\vec{p})$ leads to the appearance of power-law tails in the MD.

After integration over momentum of $f_{\gamma}(E,\vec{p})$, it is easy to obtain that the energy distribution 
$$
n_E(E)=\int f_{\gamma}(E,\vec{p}) d^3 p
\eqno{(6)}
$$
remains exponential [6]. The difference between the two distributions, Eq.(3)
and Eq.(6) is related to the quantum uncertainty.

Refs.[6,7] have been suggested to average reaction
 rates over $n(\vec{p})$ rather than distribution over energy $n_E(E)$ (see also Ref.[8]). Since (i) in this case quantum tails might produce dramatic effects
 on the rates of nuclear and other reactions in a medium and (ii) the 
provocative  suggestion [6,7] has been used in a many papers [6,7,9-18],
it is the purpose of present work to examine the validity  
of the  approximation [6,7]. 

We  stress here that directly measured cross sections of low-energy nuclear reactions in a medium, when available, are higher than the expected 
values [19-32]. To date, the observed enhancement factors are not completely
 understood. An example of the effort where environment considerations have
 been carried out for low-energy processes are the Refs.[33-35] (see also Refs.[36,37] where the tunneling of bound two-body systems through a potential 
barrier have been considered).

Let us first study the effect of the binding of the deuteron inside the 
hydrogen atom on the the example of the $p-d$ fusion reaction. The bound deuteron does not have a definite velocity. From its wave function one can calculate the deuteron velocity 
distribution 
$$
n(\vec{v}_d)=\frac{8}{\pi^2 v_0^3} \frac{1}{(1+(v_d/v_0)^2)^4},
 \eqno{(7)}
$$
where $v_0=(m_e/m_d) e^2/\hbar$, and $m_e$ and $m_d$ are the electron mass and
 deuteron mas, respectively. One has
$$
\int n(\vec{v}_d) d^3 v_d=1.
 \eqno{(8)}
$$
Now, the bound deuteron and the proton have a relative velocity
$$
v_{rel}=|\vec{v}_d-\vec{v}_p|
 \eqno{(9)}
$$
and $<\sigma v>$ is then
$$
<\sigma v>=\int S(E) \frac{1}{E} \exp(-\pi \sqrt{\frac{E_G}{E}}) v_{rel} 
n(\vec{v}_d) d^3 v_d,
 \eqno{(10)}
$$
where $E=\mu v_{rel}^2/2$, $\mu$ is the reduced proton deuteron mass, $S(E)$
is the astrophysical $S$-factor and $E_G=2 e^4 \mu/\hbar^2$.

In the case of small $v_p$ ($v_p<v_0$) the $<\sigma v>$ value, Eq.(10)
corresponds to the screening energy $E_{scr}\approx 300$eV. This result is not
 in agreement  with three body adiabatic
 calculations [8,38] (more than 10 times larger).
 In our future work we will consider an application of the Sturm-function method in the formalism of
 the three-body Faddeev-Hahn equation [38,39] for this problem.

We consider a system of $N$ identical  quantum particles carrying the unit
 positive charge, $e$, and  
contained in a volume $\Omega$ (periodic boundary conditions) with an  uniform 
external background field of the opposite sign which neutralizes the total 
charge of the  system. The time independent $n$-th state wave functions of 
the system
$\Phi_n(\vec{r}_1, \vec{r}_2, ... \vec{r}_N)$ with the energy $E_n$ 
are assumed to be normalized.

 At a thermal equilibrium, $n(\vec{p})$ can be written as
$$
n(\vec{p}) = \frac{\sum_n f_n(\vec{p}) e^{-E_n/(k_B T)}}{\sum_ne^{-E_n/(k_B T)}},
\eqno{(11)}
$$
where $f_n(\vec{p})$ is the probability to find a particle with the momentum $p$ in the $n$-th state $\Phi_n$
$$
f_n(\vec{p})= \frac{1}{(2 \pi)^3} \int e^{\frac{i}{\hbar} \vec{p}
(\vec{r}-\vec{r}^{\prime})} \Phi^{\ast}_n(\vec{r}, \vec{r}_2, ... \vec{r}_N) 
\Phi_n(\vec{r}^{\prime}, \vec{r}_2, ... \vec{r}_N)
d^3 r d^3 r^{\prime} 
\prod_{i=2}^N d^3r_i.
\eqno{(12)}
$$
A generalization of   the Kimball method, Ref.[1], leads to the following
 large-p behavior of $f_n(\vec{p})$
$$
\lim_{p\rightarrow \infty} f_n(\vec{p}) = |\psi_n(0)|^2 \frac{2}{\pi} 
\frac{N-1}{\Omega} \hbar \frac{m^2 e^4}{p^8},
\eqno{(13)}
$$
where 
$$
|\psi_n(0)|^2= \Omega \int |\Phi_n(\vec{r}, \vec{r}, \vec{r}_3,...
 \vec{r}_N)|^2d^3r \prod_{i=3}^N d^3r_i
\eqno{(14)}  
$$
and $m$ is the  particle mass. 
Substituting the large $p$ asymptotic, Eq.(13), into Eq.(11) we find the 
 large momentum tale of the momentum distribution 
 $n(\vec{p})$ at  the temperature $T$
$$
\lim_{p\rightarrow \infty} n(\vec{p}) =
\frac{2}{\pi} \hbar \rho \frac{m^2 e^4}{p^8} |\Psi(0)|^2,
\eqno{(15)}
$$
where
$$ |\Psi(0)|^2=
\frac{\sum_n |\psi_n(0)|^2  e^{-E_n/(k_B T)}}{\sum_ne^{-E_n/(k_B T)}}=
 \frac{\pi}{2 \hbar \rho m^2 e^4} \lim_{p\rightarrow \infty} n(\vec{p}) p^8
\eqno{(16)}
$$
is the contact probability of finding two particles at zero separation, and 
$\rho=(N-1)/\Omega$  is the density for $N>>1$.

We thus see that the large-p behavior is governed by the contact probability of finding two particles at 
short distance.
Although, the power-law tail in the momentum distribution was observed 
in many papers (see, for example [1-4]), it was shown  for the first 
time 
in Ref.[1] that for the ground state 
 coefficient of the $1/p^8$ term is proportional to the zero separation
 probability $|\psi_0(0)|^2$.

Now, we consider nuclear reaction between nuclei, i and j, under conditions that exist in stellar interiors. The nuclear cross-section, $\sigma_{ij}$ for small 
collision speed is approximated by
$$
\sigma_{ij}(E)=\frac{S_{ij}(E)}{E} \exp(-\pi \sqrt{E_G/E})
\approx \frac{S_{ij}(0)}{E} \exp(-\pi \sqrt{E_G/E}) ,
\eqno{(17)}
$$
where a slowly varying function with $E$ ($E<<E_{G}$)  $S_{ij}(E)$  is called 
the 
astrophysical $S$-factor, $E_{G}= 2\mu_{ij} Z_i^2 Z_j^2 e^4/\hbar^2$ is the
 Gamov energy, $\mu_{ij}=m_i m_j/(m_i+m_j)$ and $Z$ denote charge numbers.
The last equation (17) is practically exact - for example, the error in the
replacement $S(E)$ by $S(0)$ for the proton-proton fusion,
$p+p\rightarrow$$^2D+e^++\nu_e$ is only about 0.5 \% [41] for energies
 corresponding to temperatures in the center of the Sun.

The number of reactions between nuclei of i and j species at number densities 
$\rho_i$ and $\rho_j$ with a relative kinetic energy $E$ is calculated as
$$
R_{ij}=\frac{\rho_i \rho_j}{\delta_{ij}+1} \sigma_{ij}(E) v_{ij},
\eqno{(18)}
$$
where $v_{ij}=\sqrt{2 E/\mu_{ij}}$.

The Gamov rate is calculated by averaging Eq.(18) with the Maxwell-Boltzmann
 distribution $ f_{M-B}(E)$ at temperature $T<<E_G/k_B$
$$
f_{M-B}(E)=\frac{2}{k_bT} \sqrt{\frac{E}{\pi k_B T}} \exp(-\frac{E}{k_B T}).
\eqno{(19)}
$$
The result yields the Gamov rate [42]
$$
R_{ij}^G=\frac{\rho_i \rho_j}{\delta_{ij}+1}<\sigma_{ij}v_{ij}>_{M-B}=
\frac{\rho_i \rho_j}{\delta_{ij}+1}\int f_{M-B}(E) \sigma_{ij}(E) v_{ij} dE.
\eqno{(20)}
$$
It is possible to rewrite the rate in terms of the contact probability 
(the square of the wave function at the origin
 [43])
$$
R_{ij}(E)=\hbar S_{ij}(E) \frac{\rho_i \rho_j}{\pi (1+\delta_{ij})\mu_{ij}
  Z_i Z_je^2} |\psi_{ij}(0)|^2,
\eqno{(21)}
$$
where $\psi_{ij}$ is the Coulomb wave function with a normalization such that
$$
\int_{\Omega} |\psi_{ij}(\vec{r})|^2 d^3 r=\Omega
\eqno{(22)}
$$
over a large volume $\Omega$ [43].
Indeed, the square of the wave function at the origin then takes on value
$$
|\psi_{ij}(0)|^2=\frac{\pi \sqrt{E_G/E}}{\exp(\pi \sqrt{E_G/E})-1}.
\eqno{(23)}
$$

For the $N$-body system, Eqs.(11-16), the number of binary fusion reactions per unit time and unit volume in the n-th state
 $\Phi_n(\vec{r}_1, \vec{r}_2, ... \vec{r}_N)$  is
$$
R_n=S(0) \hbar \frac{\rho^2}{\pi m e^2} |\psi_n(0)|^2=
S(0) \frac{\rho}{2 (m e^2)^3} \lim_{p\rightarrow \infty}(p^8 f_n(\vec{p})).
\eqno{(24)}
$$
In (24) $|\psi_n(0)|^2$ and $f_n(\vec{p})$ are given by (14) and (13), 
respectively.

At the thermal equilibrium
$$
R=\frac{\sum_n R_n e^{-E_n/(k_B T)}}{\sum_ne^{-E_n/(k_B T)}}
= S(0) \frac{\rho}{2 (m e^2)^3} \lim_{p\rightarrow \infty}(p^8
 n(\vec{p})).
\eqno{(25)}
$$
Now, we want to see if the rate $R$, Eq.(25), and the rate $R_Q$, 
calculated in Ref.[13]
by averaging the $pp$-fusion reaction cross-section over
 quantum momentum distribution [3],
stand  in contradiction to each other
for the $pp$ fusion in the star. To do so we note that, in the
 Galitskii-Yakimets approximation  for $n(\vec{p})$ at 
 large momentum [3], ratio $R/R_Q$ is
$$
\frac{R}{R_Q}\approx 10^{3},
$$
where the standard solar model parameters [44] are used.

Clearly, the error in the 
averaging the fusion reaction cross-section over quantum momentum 
distribution
 is that one
has neglected the coupling between the various probability amplitudes of
velocity  which is introduced by the quantum uncertainty.
 
In conclusion, we summarize the main points of this letter.

(i) We have considered nuclear motion inside the atom and have found that for 
the $p - d$ fusion in the case of small $v_p$, the $<\sigma v>$ value is not in agreement with three body adiabatic calculations [8,38].

(ii) For the $N$-body system, Eqs.(11-16), we have found a general expression
 for calculating the  nuclear fusion  rate at thermal equilibrium, Eq.(25).

(iii) Our results do not support the idea of averaging the fusion reaction
 cross-section over the momentum distribution, postulated
in [6,7]
and used in many publications [6,7,9-18].

I thank N.J. Giordano for providing the opportunity to finish this work. 
I am also grateful to N.J. Fisch for valuable discussions and for making me
 aware of Ref.[9].

\pagebreak

{\bf References}
\vspace{8pt}

\noindent
1. J.C. Kimball,
J. Phys. A: Math. Gen. {\bf 8}, 1513 (1975). 

\noindent
2. E. Daniel and S.H. Vosko, Phys. Rev. {\bf 120}, 2041 (1960).

\noindent
3. V.M. Galitskii and V.V. Yakimets, Zh. Eksp. Teor. Fiz. {\bf 51}, 957 (1966) [Soviet. Phys. JETP {\bf 24}, 637 (1967)]. 

\noindent
4. P. Eisenberg et al., Phys. Rev. B {\bf 6}, 3671 (1972).

\noindent
5. L.P. Kadanoff and G. Baym, {\it Quantum Statistical Mechanics } (
 W.A. Benjamin, New York, 1962).

\noindent
6. N.L. Aleksandrov and A.N. Starostin, Zh. Eksp. Teor. Fiz. {\bf 113},
1661 (1998) [JETP {\bf 86}, 903 (1998)].

\noindent
7. A.N. Starostin and N.L. Aleksandrov, Phys. Plasmass {\bf 5}, 2127 (1998).  

\noindent
8. G. Fiorentini et al., Phys. Rev. C {\bf 67}, 014603 (2003). 

\noindent
9. A.V. Eletskii, A.N. Starostin, and M.D. Taran, Physics - Uspekhi {\bf48},
281 (2005).

\noindent
10. Y.E. Kim and A.L. Zubarev, Jpn. J. Appl. Phys. {\bf 45}, L 552 (2006);
Jpn. J. Appl. Phys. {\bf 46}, 1656 (2007).

\noindent
11. M. Goraddu et al.,  Eur. Phys. J. B {\bf 50}, 11 (2006). 

\noindent
12. M. Goraddu et al.,  Physica, A {\bf 340}, 490 (2004).

\noindent
13. M. Goraddu et al.,  Physica, A {\bf 340}, 496 (2004).

\noindent
14.  A.N. Starostin et al., Plasma Phys. Rep.  {\bf 31}, 123 (2005).

\noindent
15. A.N. Starostin, V.I. Savchenko and N.I. Fisch, Phys. Lett. A {\bf 274}, 
64 (2000).

\noindent
16. A.N. Starostin et al., Physica A {\bf 305}, 287 (2002). 

\noindent
17. A.N. Starostin et al., Physica A {\bf 340}, 483 (2004). 

\noindent
18. V.I. Savchenko, Phys. Plasmass {\bf 8}, 82 (2001). 

\noindent
19. S. Engstler et al., Phys. Lett. B {\bf 202}, 179 (1988).

\noindent
20. U. Greife at al., Z. Phys. A {\bf 351}, 107 (1995).

\noindent
21. J. Kasagi at al., J. Phys. Soc. Jpn. {\bf 64}, 3718 (1995).

\noindent
22. H. Yuki at al., J. Phys. G {\bf 23}, 1459 (1997).

\noindent
23. H. Yuki et al., JETP Lett. {\bf 68},823 (1998).

\noindent
24. K. Czerski et al., Europhys. Lett. {\bf 54}, 449 (2001).

\noindent
25. J. Kasagi et al., J. Phys. Soc. Jpn. {\bf 71}, 2881 (2002).

\noindent
26. F. Raiola et al., Eur. Phys. J. A {\bf 13}, 377 (2002).

\noindent
27. F. Raiola et al., Phys. Lett. B {\bf 547}, 193 (2002).

\noindent
28. J. Kasagi et al., J. Phys. Soc. Jpn. {\bf 73}, 608 (2004).

\noindent
29. F. Raiola et al., Eur. Phys. J. A {\bf 19}, 283 (2004).

\noindent
30. C. Rolfs, Prog. Theor. Phys. Suppl. {\bf 154}, 373 (2004).

\noindent
31. F. Raiola et al., J. Phys. G {\bf 31}, 1141 (2005).

\noindent
32. J. Cruz et al., Phys. Lett. B {\bf 624}, 181 (2005). 
  
\noindent
33. B.L. Altshuler et al., J. Phys. G {\bf 27}, 2345 (2001).

\noindent
34. M.Yu. Kuchiev, B.L. Altshuler and V.V. Flambaum, J. Phys. G {\bf 28}, 47 (2002).

\noindent
35. M.Yu. Kuchiev, B.L. Altshuler and V.V. Flambaum, Phys. Rev. C {\bf 70}, 047601 (2004).

\noindent
36. B.N. Zakhariev and S.N. Sokolov, Ann. Phys. {\bf 14}, 229 (1964).

\noindent
37.  V.V. Flambaum and V.G. Zelevinsky, J. Phys. G {\bf 31}, 355 (2005).

\noindent
38. L. Bracci et al., Nucl. Phys. A{\bf 513}, 316 (1990).

\noindent
39. A.L. Zubarev and M.Z. Nasirov, Sov. J. Nucl. Phys. {\bf 54},
389 (1991).

\noindent
40. A.L. Zubarev, M.Z. Nasirov and E.M. Gandyl, Sov. J. Nucl. Phys. {\bf 53},
566  (1991).

\noindent
41. J.N. Bahcall and R.M. May, Astrophysical J. {\bf 155}, 501 (1969).

\noindent
42. G. Gamov and E. Teller, Phys. Rev. {\bf 53}, 68 (1938).

\noindent
43. S. Ichimaru, Rev. Mod. Phys. {\bf 65}, 255 (1993). 

\noindent
44. J.N. Bahcall et al., Rev. Mod. Phys. {\bf54}, 767 (1982).

\noindent
\end{document}